\documentclass[a4paper]{jpconf}
\usepackage{graphicx}
\usepackage{epstopdf}
\usepackage{amsmath}
\usepackage{float}
\usepackage{lineno}
\usepackage{hyperref}

\begin{document}

\title{Two-pion and two-kaon femtoscopic correlations in Au+Au collisions 
  at $\sqrt{s_{NN}}$=200~GeV from STAR}

\author{Grigory Nigmatkulov (for the STAR Collaboration)}

\address{National Research Nuclear University MEPhI, Moscow, 115409, Russia}

\ead{ganigmatkulov@mephi.ru; nigmatkulov@gmail.com}

\begin{abstract}
  Measurement of femtoscopic correlations in heavy-ion collisions can
  provide information 
  about spatial and temporal parameters of the particle emission
  region at kinetic freeze-out. In this work we present the
  measurement of two-pion and two-kaon femtoscopic correlations in
  200~GeV Au+Au collisions at RHIC. The collision centrality and
  transverse momentum dependence of the three-dimensional radii,
  $R_{out}$, $R_{side}$ and $R_{long}$ is discussed.
\end{abstract}

\section{Introduction}

One of the main aims of relativistic heavy-ion collisions is to create
and study a new state of matter --- quark-gluon plasma (QGP). This
matter undergoes the rapid hydrodynamic expansion, followed by
hadronization and particle rescattering.
Two-particle correlations at small relative momentum, also known as
correlation femtoscopy or HBT, Hanbury-Brown and Twiss intensity
interferometry, are widely used to extract the spatial and temporal extent
of the particle-emitting source at the last stage of
relativistic heavy-ion collision evolution, kinetic
freeze-out~\cite{FemtoRev}. Usually femtoscopic analyses study the
most abundant pions, but with the datasets available at RHIC and LHC
it is possible to study the correlations of other particle species,
e.g. kaons. Kaons can provide complementary information to pions
because they are less affected by the feed-down from resonance decays,
have a smaller cross-section with the hadronic matter and contain
strange quarks. 

\section{Correlation femtoscopy}

The femtoscopic correlation function is defined as a ratio of the
conditional probability to observe two particles together divided by
the product of probabilities to observe each of the particles
separately. Experimentally, the correlation function is measured as a
ratio of a signal distribution, $A(q)$, that contains quantum
statistical (QS) correlations to a background distribution, $B(q)$, that
does not contain QS correlations:
\begin{equation}
  C(q) = A(q)/B(q) ,
\end{equation}
where $A(q)$ is the relative 4-momentum ($q$) distribution of particles from
the same event (collision), and $B(q)$ is a relative 4-momentum
distribution of
pairs where each particle is taken from different events (event-mixing
technique). The mixed events should have similar properties, e.g. collision
centrality, acceptance, etc. In order to get more information about
the particle-emitting source, the momentum difference is calculated in
the longitudinally co-moving system (LCMS), where the longitudinal
pair momentum vanishes, and is decomposed according to the Betsch-Pratt
convention ($q_{out}$, $q_{side}$, $q_{long}$)~\cite{BPcs1, BPcs2},
where the ``long'' axis points along the beam direction, ``out'' is
oriented along the pair transverse momentum direction, and ``side'' is
orthogonal to the other two.

The source radii are extracted from the correlation functions by the
standard Bowler-Sinyukov fit~\cite{Coul1, Coul2} of the
$C(q_{out}, q_{side}, q_{long})$ to separate the QS correlations and
Coulomb interaction:
\begin{equation}
  C(q_{out},q_{side},q_{long}) = N\left( 1-\lambda +\lambda
  K(q_{inv}) [1 + \exp(-R_{out}^2q_{out}^2 -
    R_{side}^2q_{side}^2 - R_{long}^2q_{long}^2)] \right) , \label{eqBowSin}
\end{equation}
where $N$ is a normalization factor, $\lambda$ represents the strength
of the correlations, and $R_{out}$, $R_{side}$, and $R_{long}$ are the
source radii in the ``out'', ``side'' and ``long'' directions,
respectively. The function $K(q_{inv})$ is the Coulomb part of
two-particle wave function integrated over the assumed spherical Gaussian
source with a fixed radius. In the current analysis, the value of this
radius is set to  5~fm. The $q_{inv}$ quantity is the invariant
4-momentum difference.

\section{Analysis details}

The femtoscopic analysis presented in this proceeding is applied to
the Au+Au $\sqrt{s_{NN}}$=200~GeV data taken by the Solenoidal Tracker
At RHIC (STAR) in 2011. STAR has uniform acceptance and full
azimuthal coverage. The main detector of STAR is a Time Projection
Chamber~(TPC)~\cite{STAR-TPC}. Particle identification was performed
using combined information from TPC and from Time of Flight
(TOF)~\cite{STAR-TOF} detectors. Particles are identified via specific
ionization energy loss, $dE/dx$, in the TPC gas and square of mass
determined by TOF. The collision centrality was estimated using charged particle
multiplicity at midrapidity ($|\eta|$$<$0.5). Only collisions
reconstructed within $\pm$30~cm from the center of TPC were used in
the analysis. In order to exclude interactions with the beam pipe, a
cut on the radial position of the vertex (defined as 
$V_{R}=\sqrt{V_x^2+V_y^2}$, where  $V_x$ and $V_y$ are the vertex
positions along the x and y directions) $<$2~cm was applied. Pion and kaon
candidates were required to originate from the collision vertex by
requiring the extrapolated distance of closest approach (DCA) to this
vertex to be less than 2~cm. In order to have high track reconstruction efficiency
and purity of identified particles, only tracks with $|\eta|$$<$1 and momentum
$0.15<p<1.45$~GeV/c were accepted for the analysis. Other track
quality cuts were also applied.

\section{Results and discussions}

\begin{figure}[h]
  \centering
  \includegraphics[width=0.9\textwidth]{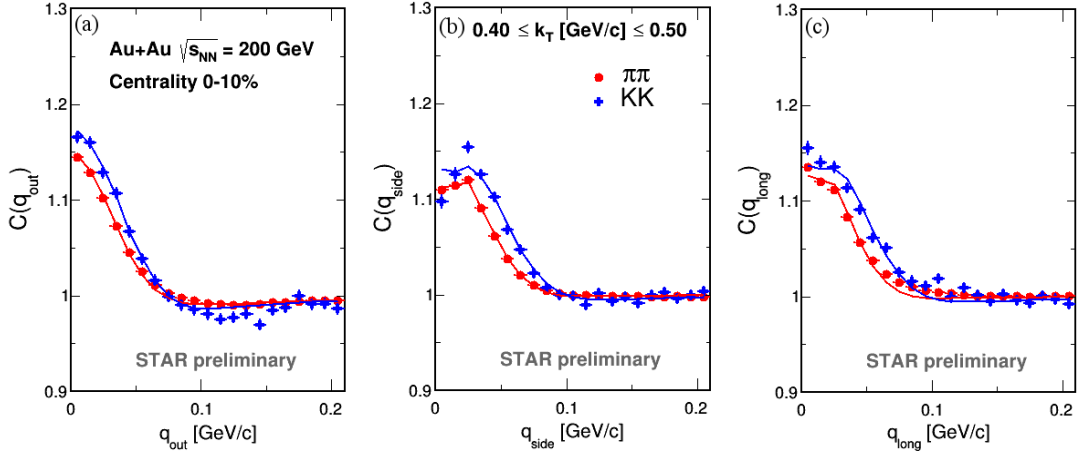}
  \caption{\label{figFits} (Color online) Sample fit projections onto
    the $q_{out}$ (a), $q_{side}$ (b) and $q_{long}$ (c) axes for
    pions (red circles) and kaons (blue crosses). The
    projections are from 0--10\% central, 200 GeV Au+Au collisions
    with $0.4<k_T<0.5$~GeV/c. Lines represent fits to the data with
    Eq.~(\ref{eqBowSin}).}
\end{figure}

Figure~\ref{figFits} shows a sample of projected $\pi^{\pm}\pi^{\pm}$
(red circles) and K$^{\pm}$K$^{\pm}$ (blue crosses) correlation
functions with fits (lines) performed according to
Eq.~\ref{eqBowSin}. Particle pairs were selected for average transverse
pair momenta $0.4<k_T<0.5$~GeV/c, where
$k_T = (p_{1}+p_{2})_{T} / 2$, and $p_1$  and $p_2$ are the three-momenta of
the first and the second particle. For the projection on one of the
directions, the relative momenta in the other two $q$ directions are
required to be less than 50~MeV/c. 

\begin{figure}[h]
  \centering
  \includegraphics[width=0.98\textwidth]{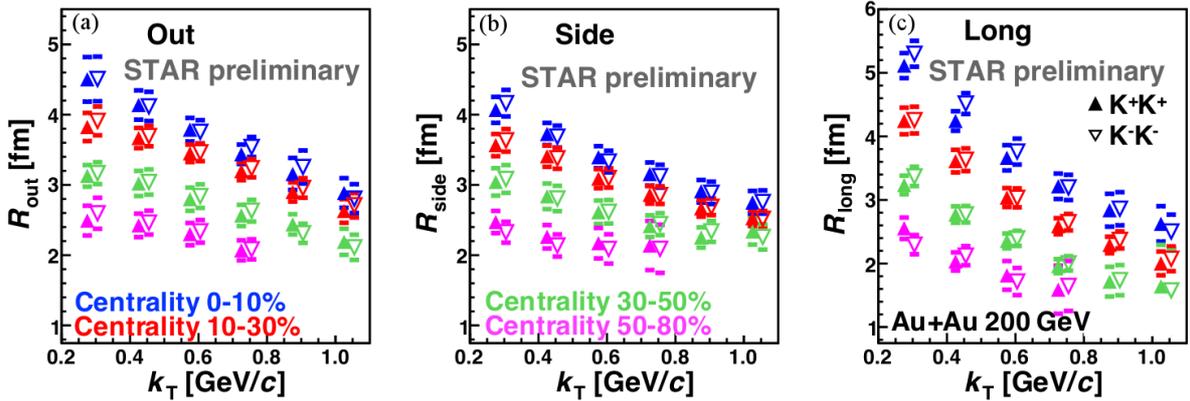}
  \caption{\label{figPos2Neg} (Color online) Transverse pair momentum
    dependencies of (a) $R_{out}$, (b) $R_{side}$, and (c) $R_{long}$
    measured for four centrality classes (0--10\%, 10--30\%, 30--50\%,
    and 50--80\%) for positive (solid triangles) and negative (open
    triangles) kaon pairs from Au+Au collisions at $\sqrt{s_{NN}}$=200~GeV.}
\end{figure}

The extracted K$^{+}$K$^{+}$ (solid triangles) and K$^{-}$K$^{-}$
(open triangles) source radii as a function of centrality and
transverse pair momentum are shown in Figure~\ref{figPos2Neg}. The
analysis was performed for 4 centrality classes (0--10\%, 10--30\%, 30--50\%,
and 50--80\%) and 6 transverse pair momentum bins. The correlation
functions for positive and negative kaon pairs were constructed
separately. It is seen that the source radii extracted for positive
and negative kaons are consistent within the uncertainties. The
decrease of $R_{out}$ and $R_{side}$ with increasing $k_T$ is an effect
of the expansion and the transverse flow. The longitudinal expansion
of the system results in the decrease of $R_{long}$ with increasing
$k_T$.

\begin{figure}[h]
  \centering
  \includegraphics[width=0.98\textwidth]{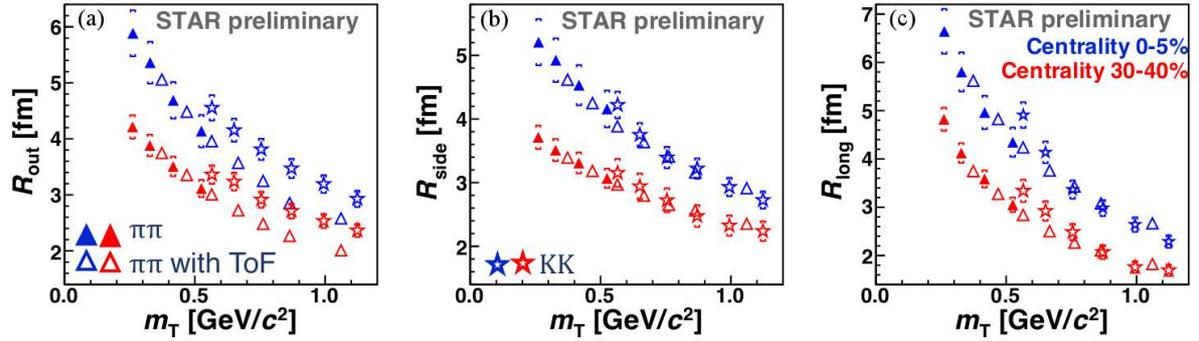}
  \caption{\label{figPi2K} (Color online) Transverse mass
    dependencies of (a) $R_{out}$, (b) $R_{side}$, and (c) $R_{long}$
    for kaons (stars) and pions (triangles) from Au+Au
    collisions at $\sqrt{s_{NN}}$=200~GeV for 0--5\% (blue symbols) and
    30--40\% (red symbols) centralities. Solid triangles represent
    pion results previously measured by STAR~\cite{STAR-PION}.}
\end{figure}

A comparison of pion and kaon source radii for 0--5\% (blue symbols)
and 30--40\% (red symbols) Au+Au collisions are shown in
Figure~\ref{figPi2K}.  The current analysis extended the previous pion
femtoscopic measurements~\cite{STAR-PION} to higher pair transverse mass
$m_T = \sqrt{k_{T}^2+m^2}$ using the TOF detector, which allows
identification of pions and kaons up to momenta $p$=1.45~GeV/c. The caps represent
systematic uncertainties for kaon and published pion
results. The systematic uncertainties for the current pion measurement are
under study. Within uncertainties, the $m_T$ dependencies of
$R_{side}$ for kaons and pions are similar suggesting the $m_T$-scaling
in the sideward direction. This may indicate that spatial extent of
pion and kaon emitting sources are similar. The $R_{out}$ values for
kaons are larger than those for pions. The $R_{long}$ for kaons and pions
have different dependence on $m_T$. Pion and kaon source radii with
similar dependences on $m_T$ as aforementioned have been reported for
Pb+Pb collisions at $\sqrt{s_{NN}}$=2.76~TeV~\cite{ALICE}.

\section{Conclusions}
Preliminary results of two-pion and two-kaon femtscopic correlations
measured in Au+Au collisions at $\sqrt{s_{NN}}$=200~GeV by the STAR
experiment have been presented. The emitting-source radii,
$R_{out}$, $R_{side}$, and $R_{long}$, are extracted from a
three-dimensional analysis. The femtoscopic radii decrease with
increasing transverse mass and decreasing charged particle
multiplicity. Qualitatively, the observed centrality and
transverse pair momentum dependencies are typical for collectively
expanding sources. Further comparisons to hydrodynamic models
are in progress.

\section*{Acknowledgments}
This work was partially supported by the Ministry of Science and
Education of the Russian Federation, grant N~3.3380.2017/4.7, and by
the National Research Nuclear University MEPhI in the framework of the
Russian Academic Excellence Project (contract No.~02.a03.21.0005,
27.08.2013).

\smallskip

\end{document}